\documentclass[usegraphicx,usenatbib,useapjfonts,apj]{emulateapj}
\usepackage{amssymb}

\def\lcdm{{$\Lambda$CDM}}
\def\mpch{\mbox{$h^{-1}$Mpc}}

\def\msunh{\mbox{$h^{-1}$M$_\odot$}}

\def\nbody{\mbox{$N-$body}}

\def\mdot{\mbox{$\dot{M}/M_{max}$}}
\def\mdotp{\mbox{($\dot{M}/M_{max}$)$_0$}}
\def\bt{\mbox{M$_{\rm lmm}/$M$_{\rm max}$}}

\def\wf{\mbox{$\tilde{\omega}_{1/2}$}}
\def\apj{\mbox{ApJ}}
\def\apjl{\mbox{ApJL}}
\def\aj{\mbox{AJ}}
\def\mnras{\mbox{MNRAS}}

\def\spose#1{\hbox to 0pt{#1\hss}}
\newcommand\lsim{\mathrel{\spose{\lower 3pt\hbox{$\mathchar"218$}}
     \raise 2.0pt\hbox{$\mathchar"13C$}}}
\newcommand\gsim{\mathrel{\spose{\lower 3pt\hbox{$\mathchar"218$}}
     \raise 2.0pt\hbox{$\mathchar"13E$}}}

\def\plotone#1{\centering \leavevmode
   \epsfxsize=\columnwidth \epsfbox{#1}}
\def\plotdelgado#1{\includegraphics[width=8.5cm]{#1}}
\def\plotancho#1{\includegraphics[width=18cm]{#1}}

\begin{document}
\slugcomment{{\em To be published in The Astrophysical Journal}}

\lefthead{DEPENDENCES OF CDM HALO ASSEMBLY ON ENVIRONMENT}
\righthead{MAULBETSCH ET AL.}

\title{The Dependence of the Mass Assembly History of Cold Dark Matter Halos on 
Environment}

\author{Christian Maulbetsch,\altaffilmark{1,2} Vladimir Avila-Reese,\altaffilmark{3}
 Pedro Col\'{\i}n,\altaffilmark{4} Stefan Gottl\"ober,\altaffilmark{1} Arman Khalatyan,
\altaffilmark{1} Matthias Steinmetz\altaffilmark{1}}

\altaffiltext{1}{Astrophysikalisches Institut Potsdam, An der Sternwarte 16, 
D-14482 Potsdam, Germany; CMaulbetsch, sgottloeber, akhalatyan, msteinmetz@aip.de}
\altaffiltext{2}{ Max-Planck-Institut f\"ur Astronomie, K\"onigstuhl 17,
D-69117, Heidelberg, Germany; maulbets@mpia-hd.mpg.de}
\altaffiltext{3}{Instituto de Astronom\'{\i}a, Universidad Nacional Aut\'onoma de M\'exico, 
A.P. 70-264, 04510, M\'exico, D.F., M\'exico; avila@astroscu.unam.mx}
\altaffiltext{4}{Centro de Radioastronom\'{\i}a y Astrof\'{\i}sica, Universidad Nacional 
Aut\'onoma de M\'exico, A.P. 72-3 (Xangari), Morelia, Michoac\'an 58089, M\'exico;
p.colin@astrosmo.unam.mx}

\keywords{cosmology:dark matter --- galaxies:formation --- galaxies:halos --- 
methods:\nbody\ simulations}

\begin{abstract} 

We show by means of a high-resolution N-body simulation 
how the mass assembly histories of galaxy--size cold dark matter (CDM) 
halos depend on environment. Halos in high density environments form 
earlier and a higher fraction of their mass is assembled in major mergers,
compared to low density environments. The distribution of the present--day
specific mass aggregation rate is 
strongly dependent on environment. 
While in low density environments only $\sim 20\%$ of the halos are not accreting 
mass at the present epoch, this fraction rises to $\sim 80\%$ at high densities.
At $z=1$ the median of the specific aggregation rate is $\sim 4$ times 
larger than at $z=0$ and almost independent on environment. All the dependences 
on environment found here are critically enhanced by local processes 
associated to subhalos because the fraction of subhalos increases as 
the environment gets denser. The distribution of the halo specific mass
aggregation rate as well as its dependence on environment resemble the 
relations for the specific star formation rate distribution of galaxies. 
An analogue of the morphology--density relation is also present at the level 
of CDM halos, being driven by the halo major merging history. 
Nevertheless, baryonic processes are necessary in order to explain
further details and the evolution of the star formation
rate--, color-- and morphology--environment relations. 

\end{abstract}

\section{Introduction}

The build--up of galactic dark matter halos is a crucial ingredient of 
galaxy formation in the context of the popular hierarchical $\Lambda$ Cold Dark 
Matter (\lcdm) scenario. Many of the present--day galaxy properties are 
expected to be tightly related to their halo mass assembly process. 
A simple way to characterize this process is by the halo 
mass--aggregation history (MAH). The MAH implies both the mass
growth by violent major mergers and by quiescent accretion.
The influence of the MAHs on the halo
 and galaxy properties remains imprinted for example on the halo concentration
\citep{AFH98, FA00, W02}, the gas infall and star 
formation rates in disk galaxies \citep[e.g.,][]{AF00,vdB02,Murali02} and 
the major merging rate and therefore, the morphology of galaxies 
\citep{KWG93,Baugh96,Rachel99,Cole00,Springel01,Granato01,
Stein02, Maller05}. 

The continuous mass growth of isolated halos is a generic process in the 
hierarchical CDM scenario \citep[e.g.,][]{Gunn82,KWG93,LC93}.
Semi--analytic models and numerical simulations have been 
used to predict the MAHs of halos of a given present--day mass,
starting from a primordial Gaussian density fluctuation field characterized 
by a power spectrum. 
On one hand, the results have shown that, due to the stochastic nature of 
the primordial density field, the MAHs of halos of a given mass at $z=0$  
span a wide range of ``tracks'', influencing galaxy features
such as the scatter of the Tully--Fisher, halo concentration--mass 
and color--magnitude relations \citep{EL96,AFH98,FA00,Bullock01,Eke01,Berlind05}. 
On the other hand, the MAHs depend on mass: less massive halos tend 
to build up a given fraction of their present--day mass 
on average earlier than 
the massive ones. This dependence is at the basis of the halo 
concentration--mass relation \citep{NFW97,FA00,Eke01,vdB02,W02,zhao03b}.

To study the general behavior of the MAHs as a function 
of mass and environment, we introduce the 
average MAH (AMAH) as $\tilde{M}(a)\equiv \frac {<M(a)>}{M_0}$, where
$a$ is the scale factor and $<M(a)>$ is the average mass of 
the most massive progenitors (MMP) of present--day halos of mass $M_0$ at 
the epoch $a$.  
It was found that the smooth AMAHs can be approximated by 
simple (universal) functions, where (1) the main function parameter is 
related to some typical formation epoch of the halo, and (2) this epoch 
depends on $M_0$ \citep{Bullock01,vdB02, W02}. 
On the basis of N--body cosmological simulations, it was also
suggested that all 
the MAHs, independent of $M_0$, present an early phase of fast mass 
aggregation (mainly by major mergers) and a late phase of slow aggregation 
(mainly by smooth mass accretion) \citep{zhao03a,Sole05,Lin05}. 
These authors also show that the depth of 
the potential well of a present--day halo, 
traced by its virial velocity, is set mainly at the end of the fast, 
major--merging driven, growth phase.

The role that the environment plays on the CDM halo assembly
has not been explicitely explored in the literature. Yet,
it is well known that some galaxy properties vary as a function
of environment. Present-day galaxy properties as well as their evolution
have been studied by some authors by combining cosmological N--body
simulations and semi--analytic modeling of the baryon galaxy processes,
a technique pioneered by \citet{Kaufetal99} and \citet{Springel01}. 
Although in the papers that used this technique the
galaxy dependencies on environment are intrinsically taken into account,
the impact the MAH dependence on environment has on the galaxy
properties is not clearly established.

The main observable dependencies with environment are seen for the 
morphological mix of elliptical and spiral galaxies \citep[e.g.,][]{Dressler80,
PG84,Whitmore93,Dom02,Goto03}, color and specific star formation rate 
\citep[e.g.,][]{Ba98,Balogh_ecol04, Ko01, Tran01, Pimbblet02, Lewis02,
Blanton03, Gomez03, Kauffmann04, Hogg04, Tanaka04,Croton05}.  Besides, there 
is some evidence that these dependencies evolve, becoming weaker at 
higher redshifts \citep{Treu03,Goto04,Bell04,sm05,Postman05}. 
As mentioned above, the morphology of galaxies as well as their colors and star 
formation rates are certainly related to the assembly history of their halos. 
Astrophysical external mechanisms, acting mainly in  
high--density environments, are also important. 
{\it To disentangle the role of 
one from the other and to understand which are the drivers of the changes of galaxy 
properties with environment, it is important to explore and quantify how 
the CDM halo assembly history depends on environment.}  N--body cosmological 
simulations are required for this endeavor.

Based on N--body simulations, \citet{LK99} concluded that only the
halo mass function varies with environment. No significant 
dependence of any other halo property on environment was found by these
authors. 
\citet{Gott01} found that the major merging rate histories of CDM halos 
vary as a function of environment. More recently, \citet{AR05} found that
some properties of halos of {\it similar masses} (for example, the
mean concentration and spin parameter) actually 
vary systematically between 
voids and clusters. Also, recently, it was shown that halos of a given mass form
statistically earlier in denser regions \citep{ST04, Harker05,Gao05,W05}.
Nonetheless, the question of how the mass assembly history of CDM
halos of similar masses depend on environment has not been
yet explored in detail. Furthermore, we would like to
know to what extent this dependence is able to explain the observed dependencies of
galaxy morphology, color, star formation rate, etc. on environment.
We would also like to know if the bimodality 
of the star formation rate (SFR) and color distributions, 
found in large statistical galaxy samples,
could be accounted for, at least in part, by the physics of dark
matter halos only.

In this Paper we construct the MAHs of $\sim 4700$ halos with 
present--day masses larger than
$10^{11}\msunh$ from a $50\mpch$ box simulation with high--mass resolution: 
$m_p=7.75 \times 10^7 \msunh$ (\S 2).  We find that several quantities that 
characterize the halo MAH (the specific mass aggregation rate and merging 
mass fraction histories, formation times, etc.) 
change significantly with environment for halos of similar present--day masses 
(\S 3). We also explore possible systematical dependences of 
observational--related quantities on intermediate--scale density, and 
discuss to what extent these dependences
are able to explain the observed galaxy property--density relations (\S 4).
Finally, we highlight the main conclusions of our work (\S 5).

\section{The simulation}

We adopt a flat cosmological model with cosmological constant (\lcdm) and
the parameters 
$\Omega_{\rm m,0}=0.3$, $\Omega_{\Lambda}=0.7$, and $h=0.7$.  The initial matter 
power spectrum has been calculated 
using the numerical results of direct integration kindly provided by W. Hu,
and it was normalized to $\sigma_8=0.9$.
The study in this Paper relies on the results from a simulation of box 
size $50\mpch$ run with the GADGET-II code \citep{Springel05}. With $512^3$ 
particles and a particle mass of  $m_p=7.75 \times 10^7 \msunh$ we are able 
to determine MAHs that extend from $z=0$ to redshifts as high as $6-9$ for halos more 
massive than $10^{11} \msunh$ (containing $>1300$ particles).
In $95$ timesteps of $\Delta a=0.01$ 
($\sim 100-160$ Myr), halos are 
identified by a new MPI version of 
the Bound Density Maxima algorithm originally introduced by 
\citet{Klypin99}. This algorithm allows to detect 
isolated or 'parent' halos (self-bound structures not contained within 
larger ones) as well as subhalos (self-bound structures contained within 
larger ones).
From these halo catalogues two merger trees were constructed, 
one for isolated halos alone and
one for all halos, including subhalos.
At $z=0$ there are more than $4700$ halos more massive than 
$10^{11}\msunh$.

A halo at redshift $z_1$ with $n_j(z_1)$ particles
is identified as the MMP of a halo
at $z_0\!<\!z_1$ containing $n_i(z_0)$ particles 
if at least a fraction $f_{min}=0.2$ of its particles are found 
in the progenitor and the overlap of particles $n_{ov}=n_i(z_0) \cap n_j(z_1)$
divided by $n_{max}=\max(n_i(z_0),n_j(z_1))$ is maximal.
$f_{min}$ is the only free parameter which is 
chosen to allow for major mergers and early rapid mass growth.
MAHs and merger trees only very weakly depend on this parameter.
For all progenitors which are merging in a timestep, at least half 
of their particles are required to be found in the descendant at $z_0$.
The algorithm to construct full
merger trees from simulations will be described in a forthcoming paper.
The construction of MAHs merely requires
the identification of the MMP which is straightforward.

The intermediate--scale environment associated to a given halo is defined 
as the density contrast of dark matter in a sphere of radius $R$ around
the halo center of mass, $\delta(R)\equiv \overline{\rho(R)}/\rho_{\rm bg}-1$
(with $\rho_{\rm bg}$ the background density  $\Omega_0 \rho_{crit}$). 
The qualitative conclusions are not dependent on the choice of $R$ for $2< R/\mpch <8$.
Since the dependences on environment become weaker for large $R$,
we will use here $\delta_4\!\equiv\!\delta(R\!=\!4\mpch)$.  
Note that with the $\delta_4$ criterion, the local environment is smoothed
out. For example, the value of $\delta_4$ for most subhalos contained within a 
given parent halo is the same, notwithstanding whether the subhalo is in the 
center or in the periphery of its parent halo.
We are here interested in exploring the general effects of environment
on the halo assembly process. The more local environmental effects are related 
only to subhalos and have been studied in detail previously 
\citep[e.g.,][]{kravtsov04,reed04,deLucia04}. 
In all cases the density contrast is defined at $z=0$, except for those panels
in Figures 8 and 9, where results for $z=1$ are shown
and $\delta_4$ is defined correspondingly at this redshift.

\begin{figure}
\plotdelgado{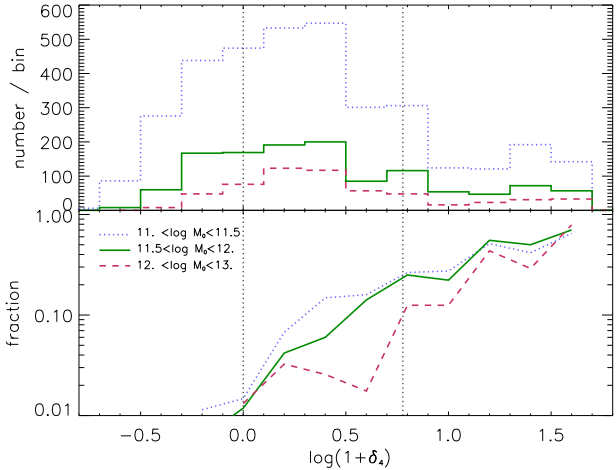}
  \caption[fig:dendis]{The upper panel shows the number distribution
of halos with density in the 3 mass ranges indicated in the plot.
The lower panel shows how the fraction of halos in the given mass range
that are subhalos depends on density contrast.
The two vertical lines indicate our cuts in density and the definition
of low density ($\delta_4<0$) and high density ($\delta_4>5$) environments. 
\label{fig:dendis}  }
\end{figure}

The distribution of the density contrast $\delta_4$ is shown in the upper
panel of Fig. \ref{fig:dendis} 
for all halos in the three mass bins used hereafter. 
The vertical lines at $\delta_4=5$ and $\delta_4=0$ illustrate 
our definitions of 'high' and 'low' density environments, respectively. 
Density contrasts $\delta_4>5$ correspond to cluster environments, 
while $\delta_4<0$ correspond to the outskirts of filaments and to voids. 
In the lower panel we show the fraction of subhalos in subsamples
selected by mass as a function of $\delta_4$. Whereas in the low density 
environment only around $1\%$ of the halos with their maximum mass 
larger than $10^{11}\msunh$ are subhalos, this fraction at high density 
amounts to about half of all halos. 
The overall fraction of halos more massive than $10^{11}\msunh$
that are subhalos rises from 7\% at
$z=1$ to 14\% at $z=0$.

We also explored the density measure used by \citet{LK99}, which 
excludes the mass of the halo itself and the close neighborhood, and 
uses only matter in a spherical shell of inner radius $2\mpch$ and
outer radius $5\mpch$ to calculate the density contrast $\delta_{2-5}$.
The results presented in the next section virtually do not change when
using $\delta_{2-5}$ instead of $\delta_4$. It should be noted that, 
according to hybrid N--body/semi--analytical models, the galaxy number 
density is proportional to the mass density \citep[e.g.,][]{Kauffmann97}.

\section{Results}

\subsection{Mass aggregation histories}

\begin{figure*}
\plotancho{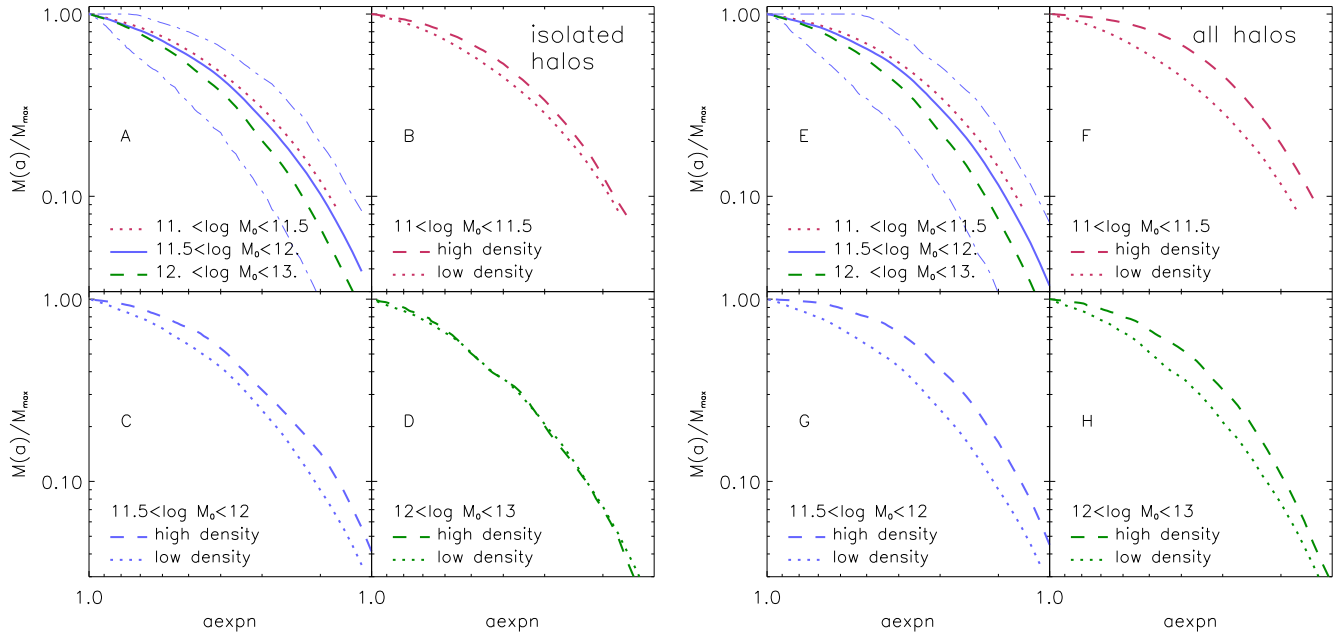}
  \caption[fig:avmahs]{Average MAHs for isolated halos are shown 
in the left 4 panels whereas AMAHs in the right 4
panels are built from all halos, including subhalos. 
The halo population is divided into 3 mass bins 
as indicated in the plot. In panels A and E the AMAHs
for all halos in these mass bins are compared.
The $10$ and $90$ 
percentiles for the middle mass bin are shown as dot-dashed lines.  
From the halo samples in the three mass bins 
two subsamples are selected at low and high density, respectively.
These are shown with mass increasing from
panel B to D and from panel F to H.
\label{fig:avmahs}}
\end{figure*}

In Fig. \ref{fig:avmahs} we present and compare the dependences of the AMAH 
on both mass and environment ($\delta_4$). The AMAHs of isolated 
halos (left four panels) and all halos (isolated+subhalos, right four panels) 
are shown in panels A and E for three different mass ranges: 
$11\!\le\!\log M_0/\msunh\!<\!11.5$,  $11.5\!\le\!\log M_0/\msunh\!<\!12$ 
and $12\!\le\!\log M_0/\msunh\!<\!13$.
Further, the AMAHs for the three mass bins, each one divided in our two 
extreme density contrasts are shown. Dashed and dotted curves correspond to the
AMAHs in high-- ($\delta_4 > 5$) and low-- ($\delta_4 < 0$) density environments,
respectively. These density--dependent AMAHs are shown with mass increasing from 
panel B to D for isolated halos, and from panel F to H for all halos.

For the construction of the AMAHs we adopt the following definitions.
Individual MAHs are normalized to their maximal mass $M_{i,max}$.
For most isolated halos the maximum mass is reached at the present day epoch, 
$M_{i,max}=M_{i}(z=0)$. If a halo suffers mass loss, 
we keep $M_i(a)$ fixed until its mass further grows. Therefore all AMAHs have 
$\tilde{M}(a\!=\!1)\!=\!1$. subhalos actually can experience substantial mass 
loss once they fall into their parent halo. 
However, it is reasonable to assume 
that the mass of a galaxy living 
in a subhalo is proportional to the maximum mass 
reached by the subhalo rather than to 
the current stripped halo mass: 
gas accretion related to the halo mass growth is stopped after the halo
falls into a host halo; 
furthermore, halo tidal stripping is not expected to 
substantially affect the mass of the galaxy
formed at the core of the halo. Therefore 
the mass of a halo
is kept fixed to the mass the halo had 
at the time it becomes a subhalo (M$_{\rm max}$),
except for those halos that suffer mergers within the host halo, 
increasing thereby their mass.

When individual MAHs are followed to high redshift, the mass of more and more 
halos falls below the resolution limit. To avoid a bias in the AMAHs
due to this incompleteness, the AMAHs are calculated only down to timesteps 
when more than $90\%$ of individual MAHs can be followed. This is the reason 
why e.g., for the lowest mass bin, the curves already end at 
$\tilde{M} \simeq 0.1$.
Since the slope of the mass function becomes steeper in lower density 
environments \citep[][]{LK99,GottVoid}, their average 
mass in a given mass bin is lower than for halos in the same mass bin 
but in high density environments. To exclude this mass effect
when comparing different environments the high-density environment 
halos are selected in their mass bin such that the average mass
is approximately equal to the bin average mass of halos in the low-density 
environment. To achieve this we have to exclude a few of the most massive 
halos in each $\delta_4>5$ bin.

Figure \ref{fig:avmahs} shows the well known dependence of AMAHs on mass:
low mass halos on average assemble a given mass fraction earlier than
massive halos. To get an impression of the scatter in individual MAHs, 
the $10$ and $90$ percentiles of the distribution are shown by dot-dashed 
lines for the middle mass bin ($11.5\!\le\!\log M_0/\msunh\!<\!12$). This 
scatter is much bigger than the differences between AMAHs 
corresponding to the different mass bins, a fact first noted by \citet[]{AFH98}.

The new result presented here is the dependence of the average halo 
MAHs of {\it similar masses} on environment. The lower is the density 
contrast $\delta_4$, the later on average
halos of similar final mass accumulate their 
mass. This ambiental dependence is already seen for 
isolated halos (left panels of Fig.\ref{fig:avmahs}).
However, the dependence becomes much 
stronger when subhalos are included, 
as shown in the right four panels of 
Fig.\ref{fig:avmahs}. 
As we saw in Fig. \ref{fig:dendis} the fraction of halos that are 
subhalos is a strong function of environment. 
On one hand, most subhalos reach their 
maximum mass at the time they fall into their host halo 
and have truncated MAHs. On the other hand, the properties of subhalos 
do not depend significantly on 
the host halo
properties or its environment \citep{deLucia04}. Thus, the main contribution of 
the MAHs of dark matter halos to the observed environmental dependence of galaxies
should result from the dependence of the subhalo fraction on
environment.

\subsection{Mass aggregation rates}

\begin{figure}
\plotone{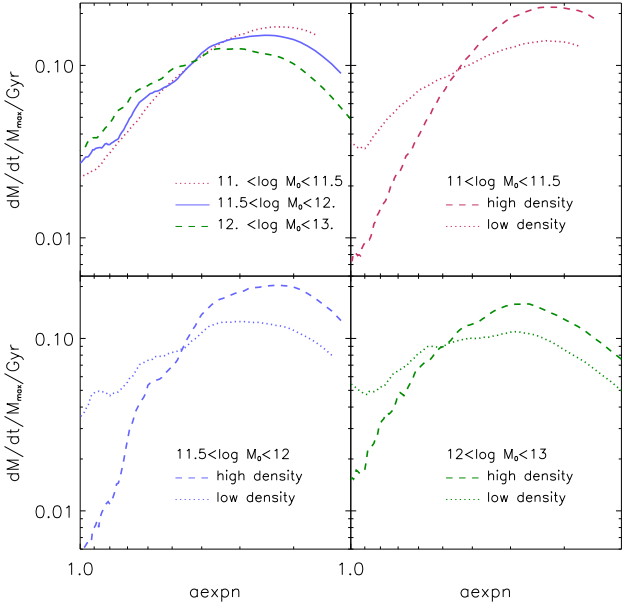}
 \caption[fig:accrate]{Evolution of the average specific mass aggregation rate 
per Gyr for halos in different mass ranges and different environments.
All (isolated+sub--) halos are included here. Line styles are as in Fig. 
\ref{fig:avmahs}.
\label{fig:accrate}}
\end{figure}

The MAHs can be further characterized by the mass aggregation rate
of the halos. We show in Fig. \ref{fig:accrate} averages of 
fractional aggregation rates per Gyr normalized to the maximum mass of the 
given halo, \mdot. In building the average \mdot, mass loss is not taken into 
account, i.e. all halos with $dM/dt<0$ have their aggregation rate set to 0 
for the reasons outlined above. 
The left upper panel compares \mdot\ for the same mass bins as in Fig. 
\ref{fig:avmahs}. While at high redshift the low mass halos are the ones 
that accumulate the mass faster, at low redshift we find that 
the high mass halos have the higher mass aggregation rates.
Thus, the maximum of \mdot\
moves slightly to lower redshifts with increasing mass. 
However, the differences in the \mdot\ histories with 
mass are not so dramatic as with environment 
(see the rest of the panels of Fig. \ref{fig:accrate}).
The average aggregations rates at $z\approx 0$ are much higher in low--density
environments than in the high--density ones, but these differences are reversed
at redshifts higher than $z\sim 1$. The former is mostly due to 
the presence of subhalos which can not accrete more mass after falling into 
their host halo (the only way to increase the mass is through mergers but 
mergers are very unlikely inside a larger virialized structure).  

\begin{figure}
\plotone{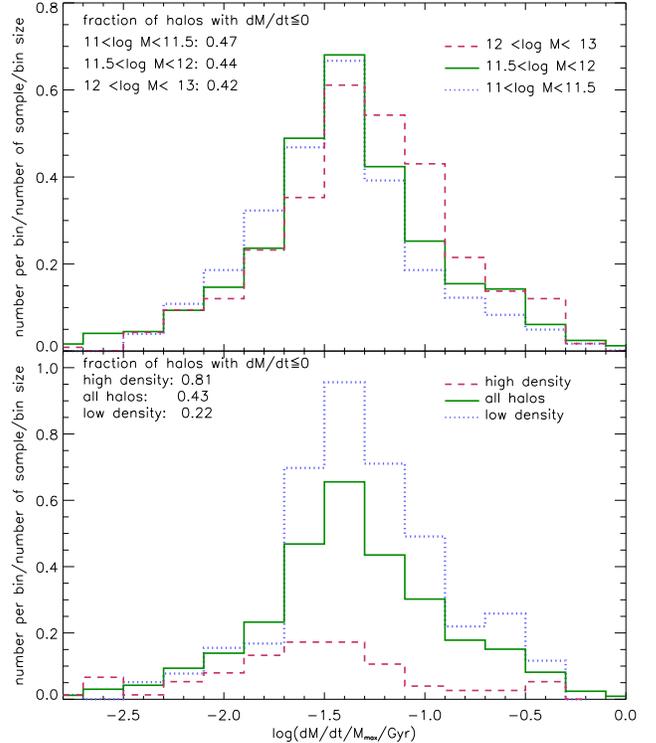}
  \caption[fig:dmaccrdis]{The distribution of the aggregation rates of dark 
matter halos between redshift $z=0.1$ and $0$ 
is shown for three different mass ranges in the upper panel
and for halos in the mass range $11.5\!\le\!\log M_0/\msunh\!<\!12.5$ 
at different density contrasts in the lower panel.
The fraction of halos with no aggregation or mass loss is indicated in the figure. 
The distribution is shown for all halos, including subhalos.
\label{fig:dmaccrdis}}
\end{figure}

The distribution of the present--day  specific mass aggregation
rate, \mdotp, consists of two parts:  43\% of 
all the halos in the mass range $11.5\!\le\!\log M_0/\msunh\!<\!12.5$ 
have $\mdotp \le 0$ ('passive' halos), while the rest has (\mdot)$_0>0$,
with a distribution 
that peaks at $\sim 0.04$Gyr$^{-1}$ (See Fig. \ref{fig:dmaccrdis}).
The distribution of (\mdot)$_0$ changes only little with mass, in the 
sense that massive halos have a slightly higher rate than the less massive 
halos (upper panel of Fig. \ref{fig:dmaccrdis}). 
On the other hand, the distribution changes dramatically with environment
(lower panel of Fig. \ref{fig:dmaccrdis}). In the mass range of 
$11.5\!\le\!\log M_0/\msunh\!<\!12.5$, the fraction of 'passive'
halos is 81\% and 22\% for the high-- ($\delta_4>5$) and low-- ($\delta_4<0$) 
density 
environments, respectively. The distribution of (\mdot)$_0$ for halos
with (\mdot)$_0 > 0$ ('active' halos) also depends on environment: the
median of the distribution is at (\mdot)$_0\approx 0.02$ Gyr$^{-1}$ and 
$\approx 0.04$ Gyr$^{-1}$ for the high-- and low-- density environments, 
respectively. 

Our results are not complete if we do not estimate the accuracy in 
the measurement of the mass aggregation rates and the fractions of passive/active 
halos. The former measurement is done by the estimate of the halo mass 
at two different output times. Therefore, its accuracy is related 
to the accuracy of the measurement of the halo mass in the simulation. 
The main source of error can be due to the halo
finder (BDM). Since the initial seeds are chosen randomly and then 
moved iteratively to the centers of candidate halos, the resulting radii 
and masses of halos can in general be different. Differences due to a 
changing particle configuration can only be detected if they are greater 
than the differences introduced by the random seeds in BDM.
We rerun the halo finder on the last timestep with different sets of initial 
seed centers for the halos. If $M_1$ and $M_2$ are two different mass 
determinations of the same halo, we find for the variance of ($M_1-M_2$)/$M_1$  
for all halos in the mass range $11.5<$log(M/\msunh)$<12.5$ the value 0.00225, 
and hence for the error in \mdotp~ a value of 0.0045. Therefore, except for 
the two lowest bins with log\mdotp$<-2.3$, Fig. 4 is not affected by these 
errors. The fraction of halos with $0<\mdotp<0.0045$ is 1.4\%.
However, we recall that the MAHs we are using in our analysis are not allowed 
to decrease with time. If we use the actual MAHs instead, we find
that $\sim 5\%$ of the halos in the above mass range have 
$-0.0045<\mdotp<0.0045$. We consider that this percentage represents 
the maximum error in the passive halo fraction determination.

Related to the mass error, there is also a systematical uncertainty in 
our estimate of the passive halo fraction. As explained and justified above, 
we used MAHs that are kept constant if the actual 
mass is falling. When using the actual mass in the two timesteps $z=0$ and 
$z=0.1$, the fraction of passive halos is 36\% instead of 43\%. On the one 
hand, the actual mass does not suffer from a temporary wrong assignment of a too 
high mass value in the past, as does the 'non-decreasing' MAHs. On the
other hand, the actual mass gives not necessarily a better estimate of the 
passive halo fraction, since the actual mass is ignoring the possibility that 
the halo could have reached its maximum mass in the past and therefore is 
passive at present. 

Summarizing, our method in 
general allows us only to find a possible range for the fractions of 
passive halos. For all the halos in the mass range $11.5<$log(M/\msunh)$<12.5$
this lies between ($36\pm 5$)\% and ($43\pm 5$)\%.

\subsection{Formation times}

Another way to characterize individual MAHs is by their
formation time.
Here we use the time when the MMP in the MAH reaches half of its maximum mass.
To isolate the ambiental effect from the known mass differential effect we use as 
variable for the formation time the scaled formation redshift, $\wf$. This 
quantity ``absorbs'' the dependence of $z_{\rm 1/2}$ on mass \citep{LC93} and 
it is defined as 
$\wf = [\delta_c(z_{\rm 1/2}) -\delta_c(z_0)] / [\sqrt{\sigma^2(M_0/2) 
-\sigma^2(M_0)}]$,
where $\delta_c(z)$ is the critical density threshold for collapse at $z$
and $\sigma^2(M)$ is the linear theory variance of density fluctuations
at mass $M$. 

\begin{figure}
\plotone{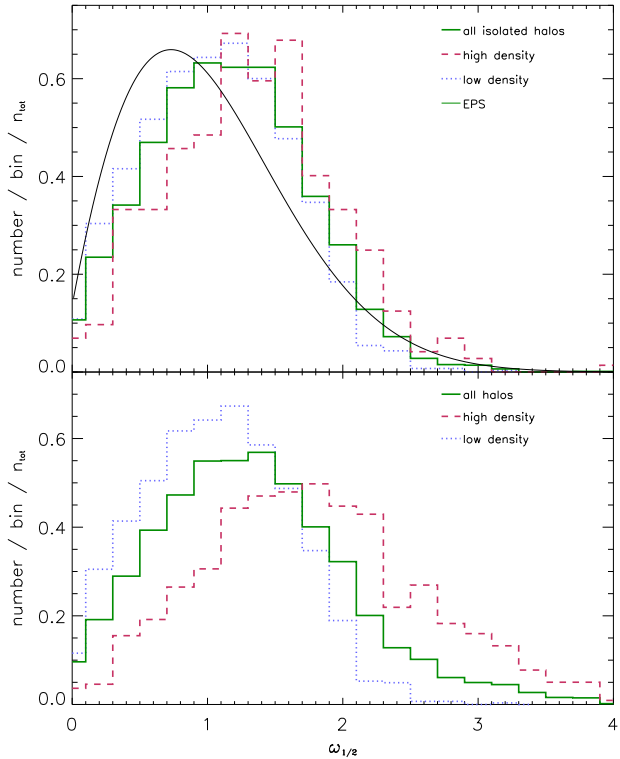}
  \caption[fig:formtime]{Distribution of formation times in terms of $\wf$.
The upper panel uses isolated halos only. The lower panel shows
the distribution of all halos, including subhalos,
in the mass range $11\!\le\!\log M_0/\msunh\!<\!13$
\label{fig:formtime}}
\end{figure}

In Fig. \ref{fig:formtime} the distribution of $\wf$ for isolated halos 
(upper panel) and all halos (isolated + subhalos, lower
panel) in the mass range $11\!\le\!\log M_0/\msunh\!<\!13$ (histograms with 
solid lines) are plotted. We further divide both samples in the two populations of 
high-density environment halos ($\delta_4>5$, histograms with dashed line),
and low-density environment halos ($\delta_4<0$, histograms with dotted line).
There is a shift in the distributions (highly significant according to a 
K-S test), the former having systematically higher scaled 
formation redshifts than the latter. Again, the difference becomes more pronounced 
when subhalos are included, because these halos 
live in the densest regions which collapse earlier and 
the mass aggregation is stopped as soon as they fall into their host halo.
The average values of $\wf$ for the sample including subhalos are $1.13\pm 0.6$ 
and $1.74\pm 0.9$ for the low and high-density environments, respectively.
This translates to formation redshifts of $z_{f,low}=1.14\pm 0.5$ and 
$z_{f,high}=1.67\pm 0.8$, respectively for $M_0=10^{12}\msunh$. In cosmic
time, this difference corresponds to $\sim 1.5$ Gyr, suggesting that galaxies
formed in the high--density environment will be redder than those formed in 
the low--density one.

For reference we also show in Fig. \ref{fig:formtime} the prediction of the 
extended Press-Schechter (EPS) 
theory \citep{LC93} using the $\Lambda$CDM power spectrum. Compared to the 
simulation result for isolated halos (see solid curve, top panel),
there is a shift to later formation times in the EPS approach
as was previously noted by \citet{vdB02}.

\subsection{Major mergers}

Major mergers are believed to play a crucial role in shaping  
galaxies and leading to an environmental dependence of galaxy morphology.
We study here further the question of the halo major--merging
dependence on environment. Earlier results by 
\citet{Gott01} showed that such a dependence exists. 
We define a major merger when the relative mass increase (MMP complement), 
$\Delta M/M$, is larger than $0.2$ 
in a timestep of $\Delta a=0.01$. 
Results do not change qualitatively by 
fixing another reasonable mass fraction increase. 
We choose this criterion to be able to find mergers
at high redshifts, where $\Delta M$ can still be determined for small halos
but no secondary progenitor would be found.
However, for subhalos we have found that the number of major mergers is overestimated 
when using $\Delta M$ to define a major merger.
Since the radius and mass of a subhalo can fluctuate due to the temporary incorrect
assignment of particles to a halo, 
by using the $\Delta M/M>0.2$ criterion more than half of 
the subhalos at $z=0$ that were more massive than $10^{11.5}\msunh$ at infall,
get a merger assigned while they were subhalos. However, no secondary progenitor 
is found for these subhalos. To overcome this problem of incorrect major--merger
counting associated to the uncertain mass determination of subhalos,
we use an alternative criterion for subhalos: a major merger is counted
when $M_{2mm}/M>0.2$, where $M_{2mm}$ is the mass of the second most massive 
progenitor. Note that to meet this condition
the second most massive progenitor should have more than 800 particles
for halos with $M>10^{11.5}\msunh$. Using this criterion, now only 5\%
of all subhalos that survive until $z=0$ suffered a major merger
once they enter in a bigger halo.

With these definitions we find that halos in high--density environments suffered
on average more major mergers than their counterparts in low-density
environments.
The average number of major mergers, counted since the mass of the 
halo is $M_i(a)>0.05 M_{i,max}$, and the standard deviations are $4.8\pm 1.4$ 
and $3.9\pm 1.3$ for high-- and low-- density environments, respectively. 
The difference in the averages is highly significant according to Student's 
$t-$test. This difference is established
mainly at $z\gsim 3$. For all the sample, we measure a mean of $4.3\pm 1.4$ 
major mergers per halo, similar to the values reported 
in \citet{Lin05}. As these authors showed, the major merging statistics does 
not depend significantly on halo mass. 

\begin{figure}
\plotone{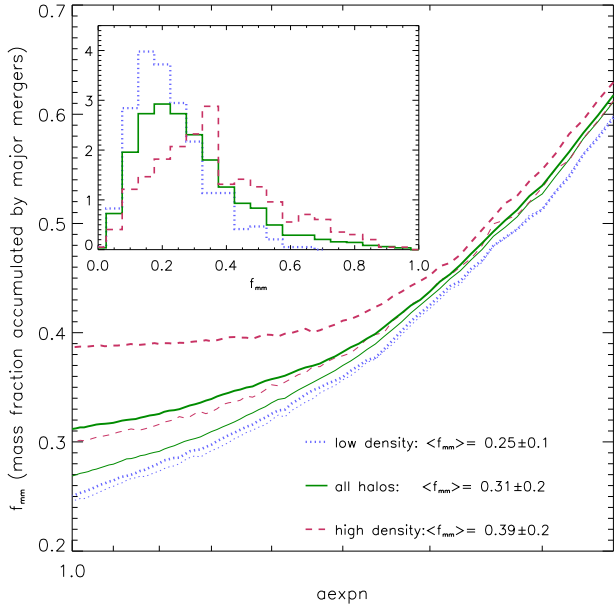}
  \caption[fig:nmm]{Evolution of the average fraction  $f_{\rm mm}$
of mass accreted in mergers ($\Delta M/M>0.2$)
is shown for all halos in the 
mass range $11.5<\log M_0/\msunh <12.5$ as solid line. Halos in
high and low density environments are shown by dashed and dotted lines,
respectively. 
The evolution of $f_{\rm mm}$ for isolated halos is indicated by thin lines.
The encapsulated panel shows the distributions of $f_{\rm mm}$
at $z=0$ for all halos. The averages and scatters of  $f_{\rm mm}$ at $z=0$
are given in the main panel.
\label{fig:nmm}}
\end{figure}

From the pure number of major mergers it is difficult
to estimate the effect mergers had on a halo, 
since early mergers only contribute a small fraction
to the final halo mass. Therefore we also measured 
for every halo the fraction of mass accreted in
major merger events. 
For the sample including subhalos and in the mass range 
$11.5\!\le\!\log M_0/\msunh\!<\!12.5$, Fig. \ref{fig:nmm} shows averages of 
the major--merger mass fraction, $f_{\rm mm}\equiv M_{\rm mm}/M$, vs $a$ 
for all halos (solid line), and for halos in the high-- 
($\delta_4 > 5$, dashed line) and low-- ($\delta_4 < 0$, dotted line) 
density environments.  At a given epoch $a$, $M$ and $M_{\rm mm}$ are 
the current halo mass and the mass assembled in major mergers until 
$a$, respectively. The encapsulated panel shows the corresponding
distribution of $f_{\rm mm}$ at $z=0$. The denser is the environment,
the higher and broader distributed is the fraction of mass assembled 
in major mergers. The fraction $f_{\rm mm}$ decreases with $a$, showing
that mass accretion becomes more and more dominant over major
mergers with time. This is particularly evident for isolated halos 
(thin lines in Fig. \ref{fig:nmm}). For the high density environment
$f_{\rm mm}$ remains constant since $z\sim 1$
because a significant fraction of halos start  
to become subhalos for which the mass is kept constant.

\begin{figure}
\plotone{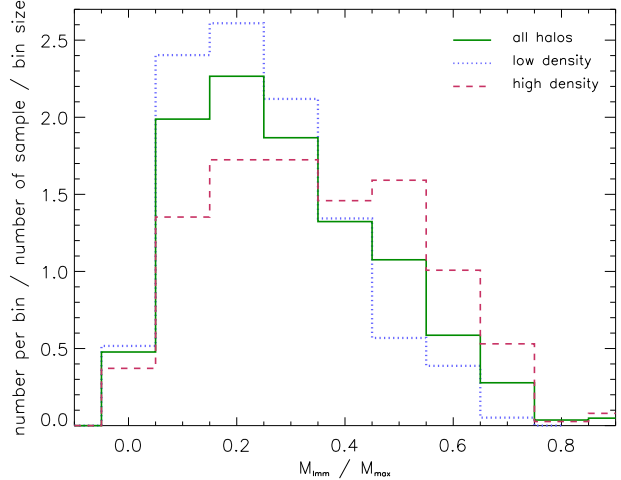}
  \caption[fig:BT]{Distribution of the ratio of the mass at the last major merger
M$_{\rm lmm}$
to the maximum mass M$_{max}$ of the halo as a proxy for the morphology of the 
galaxy living in the halo.
\label{fig:BT}}
\end{figure}

Following our results, in dense environments we expect more 
galaxies to assemble early through violent mergers. However, 
even in the densest environments, the present--day halo mass fraction aggregated 
through major/medium mergers is small (Fig. \ref{fig:nmm}). It is commonly 
assumed that the major merger of halos implies the major merger of their 
corresponding galaxies with a spheroid 
as the outcome. Further mass accretion produces again a growing disk. Then we 
may calculate the ratio of the mass when the last halo major merger happened 
to the maximum halo mass, \bt, 
in order to get a rough estimate of the spheroid-to-total mass ratio 
of the galaxy. Figure \ref{fig:BT} shows the normalized distribution of the  
\bt\ ratio for all the halos in the mass range 
$11.5\!\le\!\log M_0/\msunh\!<\!12.5$ (solid line) at $z=0$. Also shown are the
corresponding normalized distributions for the subsamples of halos
in the high-- and low--density environments.   
For high density environments the distribution is shifted to
higher \bt\ values. However, in all environments almost all halos 
still accrete at least 25\% of their mass after their last 
major merger.

\section{Discussion}

We have found that the mass assembly histories of galaxy--size CDM halos 
show marked environmental dependences, though with large scatters.
This kind of findings requires numerical simulations due to the
complexity that introduces the local and global environment. Popular
approaches used for modeling galaxy formation and evolution, as the EPS 
theory and the halo model of galaxy clustering, assume that the halo 
evolution and properties depend only on mass but not on environment or 
formation times. The results of the present paper as well as of recent 
works \citep{Gao05, Harker05, W05} should be taken into account in these 
semi--analytical approaches \citep[][]{ST04,AS05}.

In the hierarchical clustering scenario, the evolution of galaxies 
is tightly related to the assembly history of their halos. 
In this section we will discuss and attempt to quantify to what extent 
properties of the observed galaxy population are established 
already at the level of CDM halos and how the trends
with environment found for halos compare to those 
observed for galaxies.

Over the last years, a number of groups used the technique of grafting 
semi--analytic models of galaxy evolution on to CDM halo merging trees 
constructed from large--volume cosmological N--body simulations 
\citep[e.g.,][]{Kaufetal99,Diaferio01,Springel01,
Helly03,Galics03,deLucia04b,Springel05b,Kang05}. 
In these works, where the properties, distributions and evolution of  
different galaxy populations were predicted, the spatial distribution of 
galaxies is explicitely taken into account.  However, in most of these
works the halo assembly process could only be followed with accuracy 
for the most massive galaxy--sized halos and, in some cases, 
subhalos were not included in the analysis at all.
On the other hand, it was not quantified 
explicitely to what extent the obtained galaxy properties and distributions 
are the result of the CDM halo assembly history.

To tackle this question, in the following we will consider a simple model,
in which galaxy SFRs are proportional to halo aggregation rates
and bulge--to--total mass ratios (morphologies) are determined
by the \bt\ ratios of halos. Although we are well aware that this is
not a realistic model of galaxy formation, the idea is to isolate
the environmental effects of CDM and to be able to compare to observations.

\subsection{The distributions of halo properties}

The recently assembled large galaxy redshift surveys such as
the Two--degree Field Galaxy Redshift Survey
(2dFGRS) and the Sloan Digital Sky Survey (SDSS)
have revealed that specific SFRs and colors are not evenly distributed
\citep[e.g.,][and more references therein]{Strateva01,Blanton03,
Hogg04,Brinchmann03,Balogh04,Tanaka04,Kauffmann04, Weinmann06}.
Galaxies are clearly separated into two distinct populations
of red, old and 'passive' early--type galaxies, 
and blue and 'active' star forming late--type galaxies. 

The distribution of the  mass aggregation rates and its evolution
presented in \S 3.2 seem to lie at the foundation of these observed 
bimodalities of the galaxy population. While our simplifying assumption
that the specific SFR of galaxies
is driven mainly by  \mdotp\ can in general reproduce this bimodality,
the aggregation rates at the low and high--mass end 
of the halo population can not be reconciled with observations.
Here, baryonic processes must come into play.
The color distribution on the other hand is influenced not only by 
SFRs and ages of the stellar population,
but also by  the metal enrichment, i.e. a purely baryonic process.

One may consider that a fraction of the 'passive' halos 
after truncating their MAHs retain for some Gyrs a
reservoir of hot coronal gas that may still feed the galaxy with fresh gas
through cooling flows. Therefore, the fraction of galaxies formed within
the 'passive' halos that are quiescent should be lower than 
the $\approx 40\%$ of 'passive' halos found in \S 3.2.
\citet{Weinmann06} find
for the SDSS sample that roughly 31\%, 20\%, and 48\% of galaxies belong
to their categories of red quiescent, red star--forming, and blue actively 
star--forming galaxies, respectively. It is interesting to note that 
these fractions could be roughly explained at the level of CDM halo activity 
if one takes into account that the $\sim 20\%$ of red star--forming galaxies 
are formed in some of the 'passive' halos 
and in those with very low \mdotp\ values.

For the low mass halos which have low aggregation rates it is also
challenging to explain how they can host galaxies with high specific
SFRs \citep{Brinchmann03}. In these halos cooling and star formation
must be delayed with respect to the dark matter assembly.
In particular, cold gas may stay in the disk without being consumed
immediately by star formation. 
This is in general expected, as the SFRs of observed galaxies 
are related to their gas surface mass density \citep{Kennicutt98}.
For the low mass galaxies, high efficient feedback at low gas disk surface density
combined with later re-accretion of the gas could result in the high specific SFRs
at the present epoch.
The disk mass surface density is indeed predicted to be lower as the halo mass
decreases \citep[e.g.,][]{Dalcanton97,AFH98,MMW98}.

We find that the distribution of \mdotp\ is very sensitive to 
environment: 'passive' halos are the majority in the high density regions, while 
'active' halos are the majority in the low density regions 
(see fractions inside Fig. \ref{fig:dmaccrdis}). 
On the other hand, at least 
in the mass range studied in our simulation, we do 
not find a strong mass dependence in the \mdotp\ distribution (the fraction of 
halos with $\mdotp\le 0$ for three mass ranges change only by a few percent; see 
Figs. \ref{fig:accrate} and \ref{fig:dmaccrdis}).
The main reason of the environmental 
dependence of \mdotp\ 
is the fact that halos in higher density regions 
become subhalos (which means truncation of the mass aggregation process) earlier 
and much more frequently than halos in lower density regions. As a result, 
the fractions of 'passive' and 'active' halos
in the distribution of the specific mass aggregation 
rate changes significantly from low-- to high--density environment, while the 
change with mass is only marginal.

The variation of the \mdotp\  distribution with environment indeed 
resembles the  corresponding variations observed for SFR tracers 
\citep[e.g.,][see in particular Croton et al. 2005, Fig. 2,
who use an intermediate--scale environment 
criterion close to the one used here.]{Kauffmann04,Balogh_ecol04}.  
However, the shapes of the \mdotp\ distributions of halos with $\mdotp>0$ also 
change with environment. The distribution is more peaked in the low--density 
regions than in the high--density ones (Fig. \ref{fig:dmaccrdis}).
These differences are significant according to a K--S test.
The medians of \mdot\ for the halos with positive 
aggregation rates in our two selected environments are
0.02 Gyr$^{-1}$  and 0.04 Gyr$^{-1}$, respectively. 
This is at variance with the behavior 
that was reported for the population of star--forming galaxies 
by \citet{Balogh_ecol04} who found no significant differences
for these galaxies with environment.

Observations show a dependence of the bimodality in specific SFR 
or color on luminosity (or stellar mass). It is possible that 
this dependence is due to the correlation between the
galaxy luminosity function and environment \citep[e.g.,][]{Hogg03,Croton05}.
The main problem for our simplified model arises
for the massive halos in 
relatively isolated regions: they aggregate mass at a similar or slightly 
higher rate than the less massive isolated halos, in such a way  that 
the sequence of luminous red quiescent galaxies in the field is not expected; 
instead an overabundance of luminous blue active galaxies is expected. 
Observations apparently contradict these expectations: there 
is evidence of a red sequence of luminous galaxies in low density environments
and luminous blue galaxies are rare in any environment \citep{Balogh04}; 
it was also found that lower mass field galaxies tend to have higher specific
SFRs than higher mass field galaxies, and this trend continues to very high 
redshifts \citep{feulner05}. The solution to this apparent problem
may lie in astrophysical processes such as mass--dependent shock heating of the 
halo gas corona coupled to AGN feedback 
\citep[e.g.,][and references therein]{Cataneo06,Bower06}.

\subsection{Mass aggregation rate dependence on environment and evolution
of this dependence}

An interesting question with respect to the environmental dependences
is which halo property shows the strongest correlation with environment
and if some correlations are more fundamental than others.
Recent investigations \citep{Christlein05} show that the SFR--density relation 
persists, even if stellar mass, mean stellar age and morphology
are kept fixed. Also color and magnitude were shown to have the
strongest dependence on environment, compared to surface--brightness
and concentration \citep{Blanton05}.
\citet{Quintero05} even find that there is no morphology--density
relation at fixed color. 
To shed some light on the question 
of the relative strength of these effects and a possible
relation,  in Figures 8 and 9 we show 
the specific mass aggregation rates and \bt\ ratios 
(our proxies for the specific SFR and 
morphology) as a function of the density contrast $\delta_4$ and 
the cluster--centric radius, respectively.
\begin{figure}
\plotone{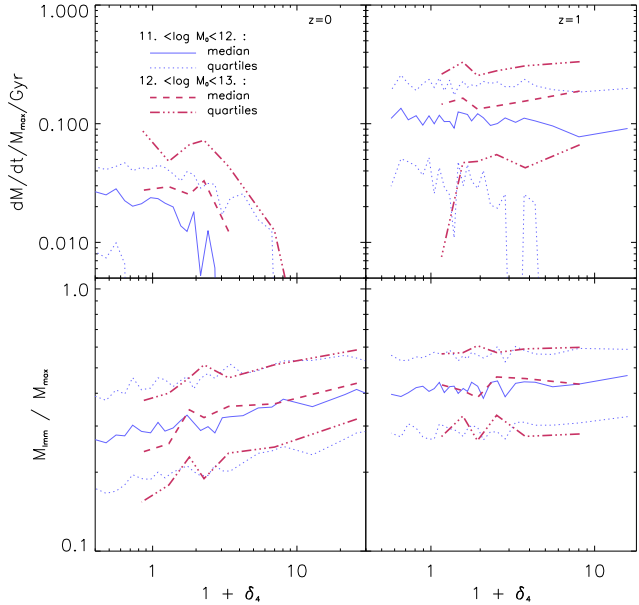}
  \caption[fig:sfrmorph]{The figure compares the density dependence
of the specific mass aggregation rates in the upper two panels
at $z=0$ and $z=1$ with the dependence of  \bt\
('morphology') on density in the lower panels.
All halos (isolated + subhalos) in the specified mass ranges
are included in the plots.
\label{fig:sfrmorph}}
\end{figure}

The overall dependence of the specific mass aggregation rate, \mdot\,
on the intermediate--scale density ($\delta_4$) 
can be appreciated in the upper panels of Fig. \ref{fig:sfrmorph}. 
Solid and dashed lines show 
the medians of all halos in the mass ranges $11<\log M_0/\msunh <12$ and 
$12<\log M_0/\msunh <13$, respectively. Dotted lines are the 25th and 75th 
percentiles for the former mass range, and the two-dotted-dashed line is the 
75th percentile for the latter mass range (the 25th percentile falls outside the 
plot).  Left panel is at $z=0$ and right panel is at $z=1$. 
The median of \mdot\ at $z=0$ anti-correlates weakly
with $\delta_4$ up to $\delta_4\approx 2-3$; for halos in higher density
environments, the median of \mdot\ drops rapidly to 0, i.e. 
the population of 'passive' halos (mostly subhalos) starts to dominate. 
The differences in these behaviors with
mass are small. The more massive halos present slightly higher  
mass aggregation rates.
At $z=1$, there is almost 
no correlation of the mass aggregation rate with environment. 
Especially halos in high density environments are still in their
growth phase and no drop of accretion rates at these densities is observed.
The  \mdot\ values at $z=1$ are systematically higher 
by a factor of  $\sim 4$ than those found at $z=0$. 
For the most massive halos,
\mdot\  even increases slightly with $\delta_4$.
The mass aggregation of the latter halos at these epochs 
happens not only by accretion but also,
in a significant fraction, by major mergers, which are common in a dense region
that still did not virialize.

Analysis of the SDSS and 2dFGRS surveys showed that SFR tracers, as the 
H$\alpha$ line equivalent width, $W($H$\alpha$), correlate with local density
of galaxies or with the cluster--centric radius 
\citep{Lewis02,Gomez03,Balogh_ecol04,Tanaka04}. The qualitative features 
of these correlations are similar to those presented here for the mass 
aggregation rate of CDM halos at $z=0$. Two populations of galaxies, those 
with significant, on--going SF, and those without SF, are revealed. 
The distributions of 
$W($H$\alpha$) or the inferred specific SFR present an
abrupt change at some characteristic density $\rho_c$ or
cluster--centric radius $r_c$: at densities higher
than $\rho_c$ (or radii smaller than $r_c$) there is a
near--total lack of star--forming galaxies, while at densities
smaller than $\rho_c$ (radii larger than $r_c$), the SF activity
tends to increase. These features are well explained if the current SFR
of galaxies is associated to the mass infall rate of their halos.

Recent studies show that the galaxy SFR and color bimodality is present at
redshifts as high as $\sim 1$, 
though with significant changes \citep{Bell04,Nuijten05,
Giallongo05,Faber05,Cooper06,Cucciati06}.  Observations show that the density of 
the galaxy red sequence roughly duplicates since $z\sim 1$ to $z=0$ with a slight
reddening of the peak of the distribution \citep{Bell04,Nuijten05}.
From our simulation, we find that the fraction of 'passive' halos with 
$\mdot\le 0$ increases by factor of $\sim 3$ since $z\sim 1$ to $z\sim 0$.

\begin{figure}
\plotone{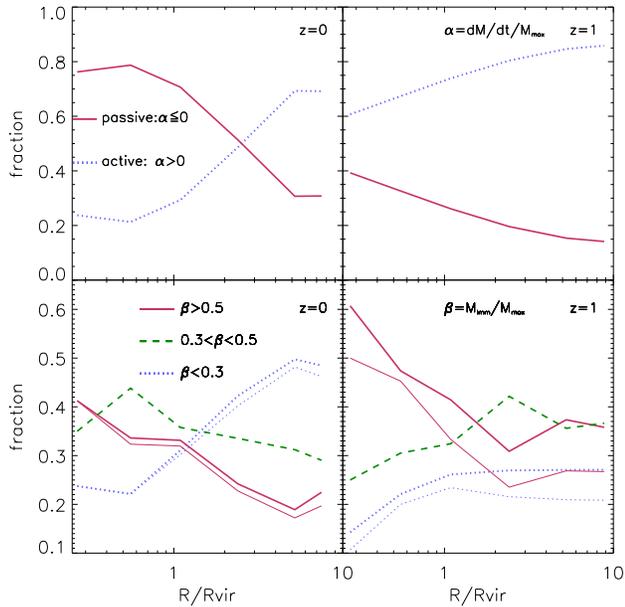}
  \caption[fig:morsfrfrac]{The upper two panels show the fractions of 'passive' 
(no accretion) and 'active' (accreting) halos 
at redshifts 0 and 1 as a function of cluster--centric radius.
The \bt\ ratio is symbolized with $\beta$.
The lower two panels show fractions of merger dominated halos 
($\beta>0.5$) as solid lines, of accretion dominated halos ($\beta<0.3$)
as dotted lines and of intermediate halos ($0.3<\beta<0.5$) as
dashed lines. Requiring that accretion dominated halos
had no major merger since $z=1$ for halos at $z=0$ or in the last 2 Gyrs
for halos at $z=1$ leads to the thin dotted lines. The additional
condition that the last major merger happened more than 2 Gyrs ago
for the merger dominated halos leads to the thin solid lines. 
\label{fig:morsfrfrac}}
\end{figure}

To further contrast 'active' and 'passive' halos,
in the upper panels of
Fig. \ref{fig:morsfrfrac} we show 
the {\it fractions} of 'active' ($\mdot >0$, dotted line) 
and 'passive' ($\mdot\le 0$, solid line) halos in our simulation
at $z=0$ (left panel) and $z=1$ (right panel)
as a function of cluster--centric radius. 
The use of cluster--centric radius allows to probe also
the radial dependence inside the group/cluster halo where 
$\delta_4$ remains constant.
At $z=0$ 'passive' halos dominate completely inside the virial 
radius of collapsed structures, $R_{vir}$.
The fraction of 'passive' halos decreases beyond $R_{vir}$ 
while the fraction of 'active' halos increases. 
In regions as far as $\sim 3R_{vir}$,
the fraction 
of 'active' halos already dominates over that one of 'passive' halos. At $z=1$ 
(i) the dependences of the fractions of both halo populations on environment
become much flatter than at $z=0$, and (ii) 'active' halos dominate over
'passive' ones even inside the virial radii of collapsed structures. 
In a very
recent work, \citet{Cucciati06} have reported a qualitatively similar evolution
for the color--density relation by using a sample of thousands of galaxies
from the VIMOS-VLT Deep Survey. We remark that the density criterion used
by these authors traces the intermediate--scale environment ($\sim$8\mpch) 
rather than the local one. 

\subsection{The morphology--density relation and its evolution}

The \bt\ ratio distributions for all (isolated + sub--) halos in the mass range 
$11.5\!\le\!\log M_0/\msunh\!<\!12.5$, and for the subsamples 
of halos in the high-- and low--density environments were presented
in Fig. \ref{fig:BT}.  The \bt~ ratios are systematically larger 
for higher environmental densities, similar to the observational trends
for the galaxy bulge--to--total luminosity ratio.
However, the fraction of halos with large \bt~  (thought to
be associated to elliptical galaxies) is very small, even in the high-density 
(cluster) environment. This suggests \citep[e.g.,][]{Maller05} that 
astrophysical processes able to 
prevent further gas accretion onto big spheroids are necessary to explain the 
observed fractions of elliptical and S0 galaxies (see above).

The bottom panels of Fig. \ref{fig:sfrmorph} show the median 
and quartiles of the \bt\ distribution
as a function of $\delta_4$ at $z=0$ (left) and 
$z=1$ (right). The line code is as in the top panels. At $z=0$, there is 
a weak but continuous increase of the \bt\ ratio with $\delta_4$.
The correlation is nearly independent of mass (if any, massive
halos have slightly larger \bt~ ratios than the less massive
halos) and it has roughly the same slope
for different percentiles. Unlike the \mdot\ distribution, there is no 
sharp change in the values of \bt\ when passing from high-- to low--density
regions. In other words, the halo \bt--density relation extends smoothly 
from high to low densities. In the case of observed galaxies, the
bulge--to--total mass ratio (morphology) indeed changes with environment more 
smoothly than the SFR \citep{Christlein05}. However, the morphology 
(mostly of faint galaxies) changes more abruptly at a certain density 
(typical of clusters outskirts) than our halo \bt\ ratio 
\citep{Tanaka04}, suggesting that baryonic processes should 
play also a role here.

At $z=1$, the dependence of the current \bt-ratio on
intermediate--scale density is almost absent 
(see lower right panel of Fig. \ref{fig:sfrmorph}). 
This plot can be interpreted as follows: in dense
environments (protoclusters), the \bt\ ratio is practically established at these
early epochs, while in less dense environments, the halos still grow
by mass accretion so that the \bt\ ratios decrease with time. 
The galaxies
inside these halos will grow still blue disks. However, some baryonic mechanisms 
that are effective in the field or poor groups can reduce
the decrease of the bulge--to--total mass ratio (and SFR) as the density is lower. 
These processes are for instance \citep{Tanaka04} the low velocity interactions 
between galaxies, able to trigger SF and consume the gas, and strangulation 
(halo--gas stripping).     

In the bottom panels of Fig. \ref{fig:morsfrfrac} the fractions of
merger dominated--halos ($\bt>0.5$), intermediate halos ($0.3<\bt<0.5$),
and accretion--dominated halos ($\bt<0.3$) are shown as a function of 
the cluster--centric--radius for redshifts $z=0$ (left panel) and 
$z=1$ (right panel). The fraction of halos with high \bt\ ratios
decreases continuously from $\sim 0.4$ in the center of virialized host halos
to $\sim 0.2$ at $4-5R_{vir}$. The fraction of halos with low \bt\ ratios
is small in the centers of virialized host halos ($\sim 0.25$), but it 
continuously increases from the periphery to regions of low density where
it attains a value of $\sim 0.5$. The fractions of halos with low
and high \bt\ ratios become comparable at $\sim 1R_{vir}$. Intermediate
halos are more abundant inside virialized host halos than in low--denstiy 
regions. 

The observed morphology--density \citep[e.g.,][]{Dressler80,PG84,Dressler85,Goto03} 
and morphology--cluster-centric radius \citep[e.g.,][]{Whitmore91,Dom02,Goto03} 
relations for galaxies are in qualitative agreement with the trends obtained here for 
halos. The fraction of intermediate--type galaxies (mostly S0's)
increases from the outskirts of clusters to intermediate cluster radii, 
while the fraction of late--type galaxies 
decreases. In the densest regions (cluster centers), the intermediate-- and 
late--type galaxy fractions decrease, while the early--type fraction 
increases. In the sparse regions (far from clusters), the morphology--density
relation flattens, i.e. the fractions of different galaxy types
only slightly change with density or cluster--centric radius, and the fraction of 
late--type galaxies dominates.  Probably, the main difference with halos 
is that the fractions of merger-- and accretion--dominated halos continue 
depending on cluster--centric radius (or on density) 
in these low--density environments 
(see Fig. \ref{fig:morsfrfrac}). However, there are several baryonic processes 
that could work in the correct direction to ``flatten'' the morphology--density 
relation (see above).

At $z=1$ the fraction of halos with high \bt\ values in the inner parts of 
clusters is higher ($\sim 0.6$) than at $z=0$ ($\sim 0.4$).  The existing 
groups/clusters at $z=1$ just formed and contain still a large fraction of halos 
that suffered a major merger very recently (large \bt\ ratios). 
At $z=0$ a large part of those halos has merged with the parent halo and is
replaced with new subhalos that had more time to aggregate mass
by smooth accretion before they became subhalos,
resulting in a higher fraction of halos with smaller \bt\ ratios than at 
higher redshift. In regions outside virialized
groups/clusters, most of the halos continue accreting mass until the present 
epoch, so that their \bt\ ratio decreases more and more, increasing therefore
drastically the fraction of halos with small \bt\ ratios (compare also Fig. 6).
  
When attempting to connect our results with observations of galaxy populations, 
we should have in mind that the halo last major merger may have happened 
very recently. In this case a high \bt\ value does not imply 
a residing early--type galaxy, but rather a star--bursting one. If we impose
the extra condition that the last major merger happened more than 2 Gyrs ago,
then the fraction of merger--dominated halos decreases, mainly 
at $z=1$ (thin solid lines in lower panels of Fig. \ref{fig:morsfrfrac}). 
Nevertheless, this fraction is still larger
at $z=1$ than at $z=0$, contrary to the observational inferences 
\citep{sm05,Postman05}. Since the ``morphological'' fractions 
reported here for halos at $z=0$ are comparable to those observed for 
galaxies, the main disagreement actually occurs at $z=1$.

On one hand, at high redshifts and in dense environments, 
most of the halos are still 
actively growing, with a significant mass fraction aggregated in mergers. 
Therefore, a part of the high--density halo population with 
$\bt>0.5$ at $z=1$ is expected to harbour actually blue, star--forming 
galaxies.
On the other hand, 
the strong increase of the fraction of accretion--dominated halos since $z=1$ 
to $z=0$, especially at large cluster--centric radii,  contrasts with the 
corresponding decrease in the observed late--type galaxy fraction 
\citep{sm05,Postman05}.
As discussed above, baryonic processes in the field and poor groups may work
to decrease the present--day fraction of late--type galaxies in the low--density 
accretion--dominated halos. However, it is not easy to explain the
observed high fraction of late--type galaxies in low--density environments 
at high redshifts as compared to the low fraction of 
accretion--dominated halos in these environments. 
Selection effects could be affecting the
observational inferences of the morphological mix.
Detailed galaxy modeling inside the evolving \lcdm\ halos is necessary
to understand better the evolution of the morphology --cluster-centric radius
and --density relations.

Finally, by comparing the strenght of the environmental dependence of the 
aggregation rate and the \bt\ ratio presented in Figures 8 and 9, we conclude
that \bt\ shows a much weaker dependence on environment than the 
aggregation rate, \mdot. 
It could be, however, that the difference between the strength
of the observed morphology--density relation and our \bt\ --density
relation is caused by processes influencing both, SFRs and morphology,
at the same time.

\section{Conclusions}

In previous works it was shown that the present--day mass function 
\citep{LK99} and internal properties \citep{AR05} of  \lcdm\ halos change 
systematically with environment. In the present paper
we investigated environmental dependences of the mass assembly process of 
\lcdm\ halos. Our results suggest
that a significant part of the relations observed between galaxy properties
and environment can be explained by the environmental dependences of the halo
assembling process. 
Our main findings are as follows:

$\bullet$  The MAHs of the \lcdm\ halos change with intermediate--scale 
environment: the lower the density contrast $\delta_4$, the later on 
average a given fraction of the maximum halo mass is assembled. The 
averages of the specific present--day mass aggregation rate, \mdotp, are 
4--5 times higher for halos in low--density environments
($\delta_4<0$) than for halos in high--density environments ($\delta_4>5$).
These differences are smaller at higher redshifts, at $z\sim 1$ disappear, and
for $z\gsim 1$ the trends are reversed. The average MAHs and mass aggregation 
rates also depend on halo mass, but to a much lesser degree than on environment. 

$\bullet$ The distribution of \mdotp\  has two parts: $\approx 40\%$ of
all the halos in the mass range $11.5\!\le\!\log M_0/\msunh\!<\!12.5$ 
do not aggregate mass, i.e (\mdot)$_0 \le 0$, while the rest
has a distribution of \mdotp\ peaked at $\sim 0.03-0.04$ Gyr$^{-1}$. 
The distribution changes little with mass but dramatically with environment: 
the fraction of 
halos with (\mdot)$_0 \le 0$ increases to $\approx 81\%$ and reduces to $\approx
22\%$ for the high-- and low--density environments, respectively.
The distribution of the halo formation time,  also changes
with environment: for instance, halos of $\sim 10^{12}\msunh$ assemble 
half of their maximum mass approximately 1.5 Gyrs earlier in the high--density 
regions than in the low--density ones.

$\bullet$ 
Using only isolated halos, the dependences of halo MAHs on environment are weak, 
showing that the intermediate--scale density around halos affects only partially 
their mass assembling process. The strongest effects of environment appear 
when subhalos are included. 
Although the effect due to subhalos is local, the
fraction of halos with masses above $10^{11}$ \msunh\ 
that become subhalos increases as the 
environment gets denser. 
The two main effects that a halo suffers
when it becomes subhalo of a larger host virialized structure are: (i) its MAH
is suddenly truncated (the mass growth is even reversed to mass loss due to 
tidal stripping), and (ii) the major merging probability drops drastically
due to the high velocity dispersions inside the virialized host halo.

$\bullet$ 
Present--day halos in high--density environments suffered more major mergers 
on average and assembled a larger fraction of their mass in major mergers than 
halos in low--density regions. 
Halos in dense regions  
become subhalos at $z\lsim 1$ much more frequently than 
those in low--density regions.
For these halos the (high) major merging mass fractions and (large) 
last major merger mass-to-current mass ratios,
 \bt,  become defined at the epoch of their infall.
On the other hand,  halos in low--density environments continue growing mainly 
by mass accretion, thus, their major merging mass fraction and \bt\ ratio 
continuously decrease.  

We have shown that the distribution of \mdotp\ and the strong dependence 
this distribution presents with environment
could explain partly the observed bimodal distributions of 
specific SFR and color for
galaxies and their systematical dependences on environment. The main difficulty 
for this direct halo--galaxy connection would be to 
explain the existence of the population of red luminous galaxies in 
low--density environments, and the high fraction of blue star forming galaxies 
among low mass galaxies. Astrophysical processes related to baryonic matter 
should be invoked here.

The morphology--density relation can be partly explained if the halo
\bt\ ratio is used as measure of the bulge--to--total mass
 ratio (morphology). However, 
even close to the cluster center, the fraction of halos with \bt\ as
high as to harbour elliptical galaxies is too small at $z=0$. This 
suggests the need of astrophysical processes able to avoid gas accretion
onto the early formed spheroids. The main trends of the observed
morphology-- density and cluster-centric radius relations at $z=0$ agree
with the trends of the halo \bt\ ratio with environment reported here.

Future comparisons of observations with models that include the 
astrophysical processes of galaxy formation in highly resolved evolving
\lcdm~ halos will tell us whether or not the \lcdm\ paradigm agrees with 
observations, and
what are the key ingredients at the basis of the morphology-, color-, and star
formation--density correlations. Our results suggest that 
the halo assembly process and its dependence on environment should
be one main ingredient.

\acknowledgments

C.M. is grateful to the Instituto de Astronom\'{\i}a at UNAM for
its kind hospitality. We acknowledge the referee for helpful
comments and suggestions.
This work has been supported by a bilateral CONACyT-DFG (Mexico-Germany) grant,
by PAPIIT-UNAM grants IN107706-3 and IN112502-3
and the DFG grant Mu-6-2.
The computer simulation discussed in this paper has been performed on the 
Sanssouci supercomputer of the AIP Potsdam.

\end{document}